\documentclass[%
reprint,
superscriptaddress,
showpacs,
 amsmath,amssymb,
 aps,
prb,
]{revtex4-1}

\usepackage{graphicx}
\usepackage{dcolumn}
\usepackage{bm}
\usepackage{color}
\usepackage[normalem]{ulem} 

\usepackage{float}

\begin{document}


\title{``Cartesian light": unconventional propagation of light in a 3D superlattice of coupled cavities within a 3D photonic band gap} 

\author{Sjoerd A. Hack}
\affiliation{%
Complex Photonic Systems (COPS), MESA+ Institute for Nanotechnology, University of Twente, P.O. Box 217, 7500 AE Enschede, The Netherlands\\}
\affiliation{%
Mathematics of Computational Science (MACS), MESA+ Institute for Nanotechnology, University of Twente, P.O. Box 217, 7500 AE Enschede, The Netherlands\\}
\email{s.a.hack@utwente.nl}

\author{Jaap J.W. van der Vegt}%
\affiliation{%
Mathematics of Computational Science (MACS), MESA+ Institute for Nanotechnology, University of Twente, P.O. Box 217, 7500 AE Enschede, The Netherlands\\}
\email{j.j.w.vandervegt@utwente.nl}

\author{Willem L. Vos}
\affiliation{%
Complex Photonic Systems (COPS), MESA+ Institute for Nanotechnology, University of Twente, P.O. Box 217, 7500 AE Enschede, The Netherlands\\}
\email{w.l.vos@utwente.nl}

\date{December 11th, 2018, in preparation for Phys. Rev. B}

\begin{abstract}
We explore the unconventional propagation of light in a three-dimensional (3D) superlattice of coupled resonant cavities in a 3D photonic band gap crystal.
Such a 3D cavity superlattice is the photonic analogue of the Anderson model for spins and electrons in the limit of zero disorder.
Using the plane-wave expansion method, we calculate the dispersion relations of the 3D cavity superlattice with the cubic inverse woodpile structure that reveal five coupled-cavity bands, typical of quadrupole-like resonances. 
For three out of five bands, we observe that the dispersion bandwidth is significantly larger in the $(k_x, k_z)$-diagonal directions than in other directions. 
To explain the directionality of the dispersion bandwidth, we employ the tight-binding method from which we derive coupling coefficients in 3D. 
For all converged coupled-cavity bands, we find that light hops predominantly in a few high-symmetry directions including the Cartesian $(x, y, z)$ directions, therefore we propose the name "\textit{Cartesian light}". 
Such 3D Cartesian hopping of light in a band gap yields propagation as superlattice Bloch modes that differ fundamentally from the conventional 3D spatially-extended Bloch wave propagation in crystals, from light tunneling through a band gap, from coupled-resonator optical waveguiding, and also from light diffusing at the edge of a gap. 
\end{abstract}

\pacs{Valid PACS appear here}
\maketitle



\section{Introduction}

Ever since the emergence of the field of nanophotonics, it is well-known that fruitful analogies can be drawn between the behavior of photons at the nanoscale on the one hand, and the physics of electrons, spins, and phonons in condensed matter on the other hand~\cite{vanHaeringen1990Book, Lagendijk1996PR, vanRossum1999RMP,Soukoulis2001Book,Akkermans2007Book,Ghulinyan2015Book}.
The seminal phenomenon considered in this respect was the three-dimensional (3D) Anderson localization of light~\cite{John1984PRL,Anderson1985PM} - in analogy to Anderson localization of spins~\cite{Anderson1958PR} - that continues to receive attention to date~\cite{vanderBeek2012PRB,Sperling2016NJP}.
Other well-known examples are the analogy between weak localization or enhanced backscattering of light~\cite{vanAlbada1985PRL,Wolf1985PRL} and of electrons~\cite{Bergmann1984PR}, and the analogy between a complete 3D photonic band gap in a 3D photonic crystal~\cite{Yablonovitch1987PRL,John1987PRL,Joannopoulos2008Book} and the electronic band gap in a semiconductor crystal such as silicon or germanium~\cite{Ashcroft1976Book}. 

In this work, we explore the propagation of light in a 3D superlattice of coupled resonant cavities inside a 3D photonic band gap.
The propagation of light in such a 3D cavity superlattice is analogous to electronic transport in an impurity band in a semiconductor~\cite{Anderson1958PR,Feher1959PR,Ashcroft1976Book}.
Light hops from cavity to cavity throughout the 3D superlattice, which differs fundamentally from the conventional spatially-extended Bloch wave propagation outside the gap.
Since the light hops predominantly in a few high-symmetry directions including the Cartesian $(x, y, z)$ directions, we propose the name "Cartesian light" for the unusual propagation of light in the 3D superlattice of coupled cavities in a 3D photonic band gap. 

In one dimension (1D), a chain of coupled resonant cavities is a well-known system that is known as a coupled resonator optical waveguide (CROW)~\cite{Yariv1999OL}.
The weak coupling between cavities in a CROW has been demonstrated at optical frequencies \cite{Notomi2008NP}.
CROWs are widely studied for efficient nonlinear optical frequency conversion and for perfect transmission through bends, and for the one-dimensional (1D) localization of light~\cite{Yariv1999OL,Mookherjea2008NatPhoton}. 
Two dimensional (2D) arrays of coupled cavities have been studied, notably 
for unusual discrete diffraction effects~\cite{Christodoulides2003N}, for intricate coupled nanolasers~\cite{Altug2005OE}, for topologically-protected propagation~\cite{Hafezi2013NP}, and for transverse localization~\cite{deRaedt1989PRL,Schwartz2007N}.
The coupling between the cavities at optical frequencies has been demonstrated to be significantly larger than the fabrication-induced disorder in the cavity frequencies~\cite{Majumdar2012PRB}.
In 3D resonator arrays without band gap, topologically-protected propagation was studied~\cite{Lin2016NatCommun,Zhang2017PhysRevA}, as well as the percolation of light through 3D lattices of coupled resonant microspheres~\cite{Astratov2007OptExpress}, and dynamic localization of light~\cite{Yuan2015PRL}. 
In 3D photonic band gap crystals in the microwave regime, slow heavy-photon propagation was reported in a 1D array of weakly-coupled cavities~\cite{Bayindir2000PRL,Bayindir2000PRB1,Bayindir2000PRB2}.
Numerical calculations of a 2D array of cavities embedded in a 3D woodpile photonic crystal revealed ultraslow and negative group velocities~\cite{Li2006OL}. 
To the best of our knowledge, 3D superlattices of coupled cavities with resonances in a 3D photonic band gap have not yet been studied before. 

\section{Methods}
\label{sec:structures}

\begin{figure}[tbp]
\includegraphics[width=1.00\columnwidth]{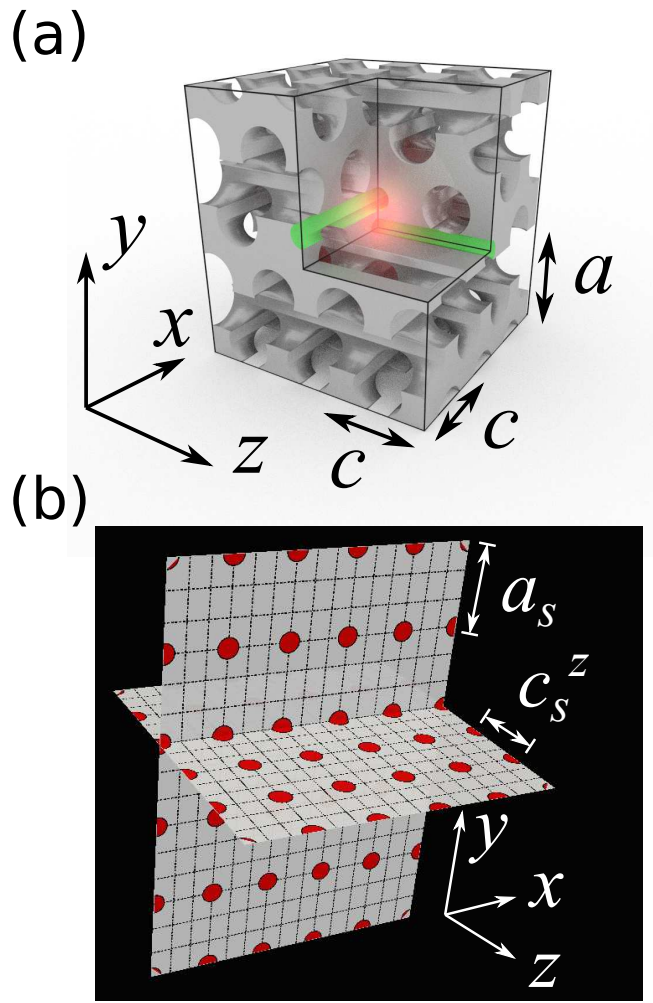}
\caption{
(a) Design of a single cavity in an inverse woodpile photonic band gap crystal shown in a cut-out of a $M_x M_y M_z = 3\times3\times3$ supercell that is surrounded by boxed lines. 
The high-index backbone is shown in gray. 
Two proximal smaller defect pores are indicated in green, and the cavity region is highlighted as the bright region at the center. 
The tetragonal lattice parameters $a$ and $c$ are shown, as well as the $x,y,z$ coordinate system. 
(b) $(x,z)$ and $(x,y)$ cross sections through a 3D superlattice of resonant cavities, with red circles indicating cavities and dashed rectangles representing unit cells of the underlying inverse woodpile crystal structure (see (a)). 
The lattice parameters ($c_s^{x}$, $a_s$, $c_s^{z}$) of the superlattice are shown, as well as the $x,y,z$ coordinate system. 
}
\label{fig:crystalOfCavities}
\end{figure}

In this paper, we study a 3D cavity superlattice that is embedded in a 3D photonic band gap crystal that has the inverse woodpile structure. 
This structure has nearly the same symmetry as a diamond crystal of carbon atoms,\cite{Ho1994SSC} yet thousandfold magnified, as illustrated in a youtube animation.\cite{COPS2012youtube} 
The inverse woodpile crystal structure consists of two perpendicular 2D arrays of nanopores with radius $r$ in a high-index medium such as silicon,\cite{Ho1994SSC} as illustrated in Figure~\ref{fig:crystalOfCavities}(a). 
Each 2D pore array corresponds to a diamond ${110}$ crystal face. 
In view of the arrangement of the nanopores, it appears to be convenient to employ a tetragonal unit cell~\cite{Hillebrand2003JAP,Woldering2009JAP} instead of the conventional cubic unit cell~\cite{Ashcroft1976Book}. 
The tetragonal unit cell has lattice parameters $c$ (in the $x$ and $z$ directions), and $a$ (in the $y$ direction) in a ratio $a/c = \sqrt{2}$ to ensure a cubic crystal structure. 
More details, notably on the Brillouin zone, are presented in Appendix~\ref{sec:BrillouinZoneTetragonalUnitCell}. 

Inverse woodpile photonic crystals possess a broad 3D photonic band gap, whose width strongly depends on the radius $r$ of the pores~\cite{Ho1994SSC,Hillebrand2003JAP,Woldering2009JAP}. 
For a normalized pore radius $r/a=0.24$ - as considered here - a maximum relative bandwidth $\Delta \omega_{pbg}/\omega_c = 25.3\%$ occurs for $\epsilon = 12.1$ typical of silicon,\cite{Hillebrand2003JAP,Woldering2009JAP} with $\Delta \omega_{pbg}$ the frequency width of the band gap, and $\omega_c$ the band gap's center frequency. 
3D inverse woodpile crystal nanostructures have been fabricated from a number of different high-refractive index backbones~\cite{Schilling2005APL, Santamaria2007AdvMater, Hermatschweiler2007AdvFunctMater, Jia2007JApplPhys, vandenBroek2012AFM, Grishina2015NT}. 
In nanophotonic experiments, the potential of silicon inverse woodpiles was demonstrated by the observation of a broad 3D photonic band gap for many angles~\cite{Huisman2011PRB}, as well as a strong spontaneous emission inhibition of embedded quantum dots~\cite{Leistikow2011PRL}.

To create a resonant cavity in a inverse woodpile photonic crystal, Ref.~\cite{Woldering2014PRB} proposed a design whereby two proximal perpendicular pores have a smaller radius $(r' < r)$ than all other pores, as shown in Figure \ref{fig:crystalOfCavities}(a). 
Near the intersection region of the two smaller pores, the light is confined in all three directions to within a mode volume as small as $V_{\text{mode}}=\lambda^3$ where $\lambda$ is the free-space wavelength.\cite{Woldering2014PRB} 
Supercell band structures revealed up to five resonances within the band gap of the perfect crystal, depending on the defect pore radius $r'$\cite{Woldering2014PRB}, that have quadrupolar symmetry.\cite{Devashish2018Arxiv} 
The best confinement occurs for a defect radius $r'/r=0.5$ that is also considered here. 

Figure~\ref{fig:crystalOfCavities}(b) shows a 3D superlattice of cavities as is studied here, where each sphere indicates one cavity, as shown in Fig.~\ref{fig:crystalOfCavities}(a).
The cavity superlattice has lattice parameters ($c_s^{x}$, $a_s$, $c_s^{z}$) in the ($x, y, z$) directions that are integer multiples of the underlying inverse woodpile lattice parameters: $c_s^{x} = M_x c$, $a_s = M_y a$, $c_s^{z} = M_z c$. 
Here, we study the $M_x M_y M_z = 3\times3\times3$ superlattice such that the cavities are repeated every three unit cells with lattice parameters $c_s^{x} = 3c$, $a_s = 3a$, $c_s^{z} = 3c$. 
Thus, the cavity superlattice is also cubic, similar to the underlying inverse woodpile structure (see section~\ref{sec:discussion} for additional discussion). 

We have calculated the band structure of the 3D cavity superlattice using the plane-wave expansion method.\cite{Ashcroft1976Book,Joannopoulos2008Book,Johnson2001OptExpress} 
Using the Richardson extrapolation method allows us to estimate the frequencies in the limit of infinite grid resolution~\cite{Roache1998Book,Richardson1927PhilosTransRoyalSocA}.  
Details on the calculations and the convergence are given in Appendix \ref{sec:Calcs_Convrg}. 
All calculations were performed on the ``Serendipity" cluster in the MACS group at the MESA$^{+}$ Institute.\cite{Serendipity} 
Even on this powerful computer cluster the calculations took 210 hours. 

\section{Results}
\label{sec:results}
\subsection{Band structure of coupled cavity resonances}

\begin{figure}[tbp]
\includegraphics[width=1.00\columnwidth]{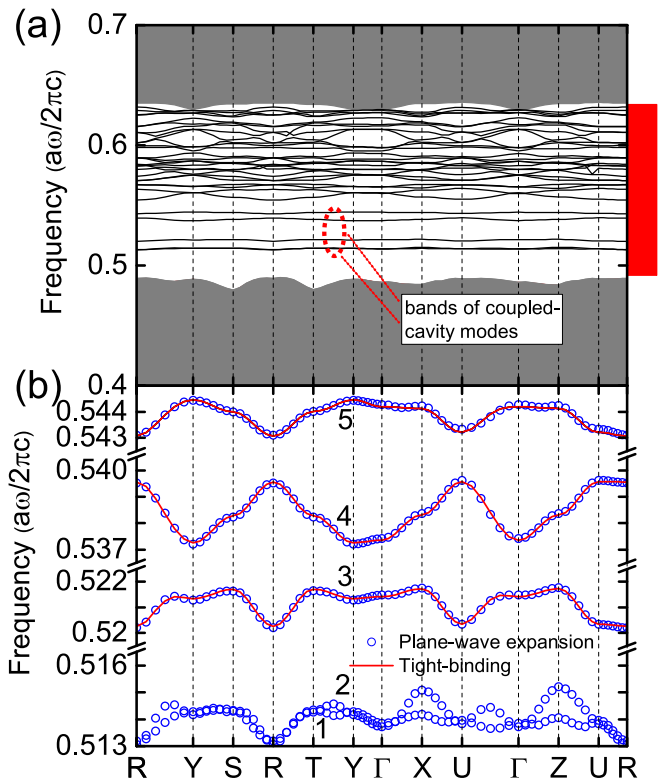}
\caption{
(a) Black curves are photonic bands of a 3D inverse woodpile photonic band gap crystal ($r/a=0.24$, $\epsilon_{\text{Si}}=12.1$) with two proximate defect pores ($r'/r=0.5$). 
The abscissa indicates the wave vector between the high-symmetry points of the Brillouin zone (see Appendix~\ref{sec:BrillouinZoneTetragonalUnitCell}). 
The 3D photonic band gap of the perfect crystal is shown as a red bar, and the range of allowed modes outside the band gap is shown in grey. 
The five bands of coupled-cavity modes are highlighted. 
(b) Zoom-in of the five coupled-cavity bands (blue circles), labeled $m = 1,\ldots, 5$. 
The $m = 3, 4, 5$ bands are accurately described by the tight-binding model (red curves). 
} 
\label{fig:bandStructureCrystalOfCavities}
\end{figure}

In Fig.~\ref{fig:bandStructureCrystalOfCavities}(a), the band structures are shown of a 3D cavity superlattice ($M_x M_y M_z = 3\times3\times3$) in an inverse woodpile photonic band gap crystal made of silicon. 
As a result of the intentional defect pores many bands appear in the band gap of the perfect crystal between reduced frequencies $\tilde{\omega} = \omega a/ (2 \pi c) = 0.492$ and $0.634$. 
The lowest five bands between $\tilde{\omega} = 0.5$ and $0.55$ are dispersionless and correspond to the five cavity resonances that are from now on labeled as $m = 1,\ldots,5$. 
The $m=3,4,5$ bands are isolated in frequency, unlike the situation in solid state physics where the bands arising from $d$-orbitals are hybridized~\cite{Ashcroft1976Book}.
The dispersions of the bands in Fig.~\ref{fig:bandStructureCrystalOfCavities}(a) agree well with those of Woldering \textit{et al.}~\cite{Woldering2014PRB}. 
The dispersive bands in the top half of the gap (between $\tilde{\omega} = 0.55$ and $0.634$) have unknown character, and may include waveguiding along the defect pores. 

A closer inspection of the five dispersionless cavity bands in Fig.~\ref{fig:bandStructureCrystalOfCavities}(b) reveals that these bands have nonzero bandwidths, indicating that cavity resonances in the $M_x M_y M_z = 3\times3\times3$ superlattice are coupled, as is investigated in this paper. 
Our results agree well with a simultaneous investigation of a single cavity in an inverse woodpile crystal with finite support, studied by other numerical methods.\cite{Devashish2018Arxiv} 
Notably, Ref.~\cite{Devashish2018Arxiv} also reports that the first two bands are nearly degenerate. 
Based on the occurrence of five cavity bands, on degeneracies between bands, and on the field distribution (reported in Ref.~\cite{Woldering2014PRB}), it has been concluded that the resonances of the inverse woodpile cavity have quadrupolar symmetry and are the optical analogues of d-orbitals in solid-state physics~\cite{Devashish2018Arxiv}. 
Therefore, it is naively expected that neighboring cavities couple in diagonal directions. 

\subsection{Dispersion bandwidths}
For a 1D coupled-resonator optical waveguide (CROW), it is well-known that the coupling coefficient along the waveguide is proportional to the dispersion bandwidth.\cite{Yariv1999OL,Bayindir2000PRB2,Altug2004ApplPhysLett} 
A straightforward extension of this notion to a 3D cavity superlattice is to consider the dispersion bandwidth in various crystal directions, since this is straightforward to derive from photonic band structures as in Figure~\ref{fig:bandStructureCrystalOfCavities}. 
For a given crystal direction characterized by wave vector $\mathbf{k}$, the dispersion bandwidth is defined as 
\begin{equation}
\Delta \tilde{\omega} \equiv |\tilde{\omega}_{max} - \tilde{\omega}_{min}|_{\Gamma\to\mathbf{k}_{BZ}}
,
\label{eq:dispersion-bandwidth}
\end{equation}
in other words, the absolute value of the difference between the maximum and minimum frequencies on a trajectory in reciprocal space between the origin $\Gamma$ and the edge of the Brillouin zone $\mathbf{k}_{BZ}$ in the direction of $\mathbf{k}$. 
As an example, for the $m = 3$ coupled-cavity band in Figure~\ref{fig:bandStructureCrystalOfCavities}(b), between $\Gamma$ and $Z$ the minimum and maximum frequencies are nearly the same ($\tilde{\omega} = 0.521$) hence the bandwidth $\Delta \tilde{\omega}$ is nearly zero. 
Between $\Gamma$ and $U$ the minimum and maximum frequencies differ much more ($\tilde{\omega} = 0.520$ to $0.521$) hence the dispersion bandwidth is much greater in the diagonal direction. 

\begin{figure}[h!]
\includegraphics[width=1.00\columnwidth]{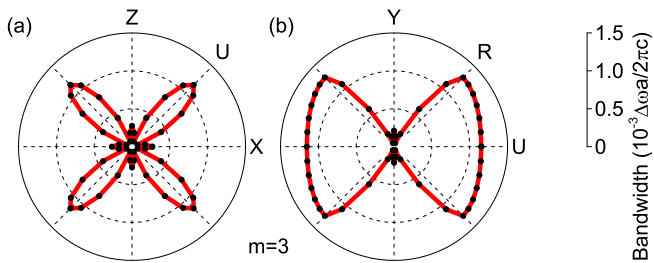}
\caption{
(a) Polar plot of the dispersion bandwidth for the $m = 3$ coupled-cavity band in the $(k_X,k_Z)$ plane. 
The $X$, $U$, and $Z$ high-symmetry points are shown. 
(b) Polar plot of the dispersion bandwidth for the $m = 3$ coupled-cavity band in the $(k_Y,k_{U})$ plane. 
The $Y$ and $U$ high-symmetry points are shown. 
The black circles indicate the plane-wave results (cf. Fig.~\ref{fig:bandStructureCrystalOfCavities}), and the red lines are guides to the eye. 
}
\label{fig:dispersionBandwidthThird}
\end{figure}

A polar plot of the dispersion bandwidth $\Delta \omega$ versus wave vector $\mathbf{k}$ in the $(k_X,k_Z)$ plane is shown in Figure~\ref{fig:dispersionBandwidthThird}(a). 
From the band frequencies mentioned above, the dispersion bandwidth is very small in the real-space $x$ and $z$-directions (corresponding to $\Gamma X$, $\Gamma Z$, respectively, in reciprocal space). 
In the diagonal directions that correspond to the $\Gamma U$ high-symmetry trajectory the dispersion bandwidth is much greater. 
As seen from a given central cavity in real space, the wave vector $\mathbf{k}$ is then directed towards a second nearest neighboring cavity in the diagonal $1/\sqrt{2}.(1,0,1)$ direction (see Appendix~\ref{sec:BrillouinZoneTetragonalUnitCell}). 
The polar plot of the dispersion bandwidth for $m = 3$ therefore looks like a quadrupolar radiation pattern. 
Based on the 1D CROW-reasoning given above, one tentatively infers that light is transported through the 3D cavity superlattice preferentially in the $xz$-diagonal (corresponding to $\Gamma U$) directions. 

Figure~\ref{fig:dispersionBandwidthThird}(b) shows the dispersion bandwidth $\Delta \omega$ in the $(k_Y,k_{U})$ plane. 
The largest bandwidth occurs at about $45^{o}$ off the $(XUZ)$ plane, which corresponds to the $\Gamma R$ high-symmetry direction. 
The bandwidth in the $\Gamma Y$-direction is small, from which one tentatively infers that there is little light transport in the $y$-direction in real space. 

\begin{figure}[h!]
\includegraphics[width=1.00\columnwidth]{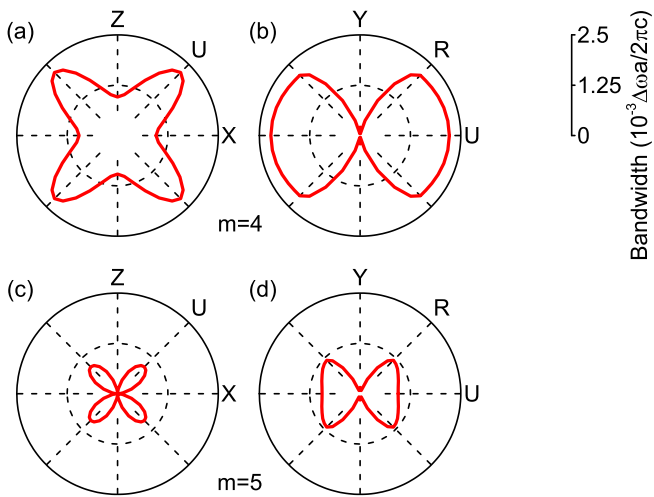}
\caption{
Polar plots of the dispersion bandwidth versus wave vector for the 
(a,b) $m = 4$ and (c,d) $m = 5$ coupled-cavity bands. 
Panels (a,c) are in the $(k_X,k_Z)$ plane and (b,d) are in the $(k_Y,k_{U})$ plane.
}
\label{fig:dispersionBandwidthFourthFifth}
\end{figure}

The dispersion bandwidth for the $m = 4,5$ bands is shown in Figures~\ref{fig:dispersionBandwidthFourthFifth}(a,b). 
For the $m = 4$ band, the dispersion bandwidth is large in the diagonal directions that correspond to the $\Gamma U$ high-symmetry directions, and it is smaller in the $x$ and $z$-directions ($\Gamma X$,$\Gamma Z$, respectively). 
Compared to the $m = 3$ band, the dispersion bandwidth for the $m = 4$ band appears to be less strongly directional. 
To quantify the directionality, we consider a directionality $D$ ratio between the maximum and the minimum bandwidths $D = \Delta \omega_{max}/\Delta \omega_{min}$ in the $(k_X,k_Z)$ plane, which yields a directionality of about $D = 4$ that is much lower than $D = 15$ for the $m = 3$ band. 

For the $m = 5$ band in Figure~\ref{fig:dispersionBandwidthFourthFifth}(c), the polar plot of the dispersion bandwidth looks very much like a quadrupolar emission pattern. 
The bandwidth is small in the diagonal directions that correspond to the $\Gamma U$ high-symmetry directions, about $2.5 \times$ smaller than for the $m = 4$ band. 
The dispersion bandwidth is much smaller in the $x$ and $z$ directions, corresponding to a large directionality $D = 26$. 

\subsection{Coupling coefficients}
To understand the coupling between the cavities in a 3D cavity superlattice more fundamentally, we derive the coupling coefficients of light from the dispersion relations using the tight-binding method, see Appendix \ref{sec:photonictightbindingmethod} for details. 
Figure~\ref{fig:bandStructureCrystalOfCavities}(b) shows that the $m = 3$ band is accurately described by the tight-binding model. 
It appears that only $7$ independent coupling coefficients $\kappa$ are needed in the tight-binding model, namely for the real-space directions $x$, $y$, $z$, $xz-diagonal$ (corresponding to $\Gamma U$ in reciprocal space), $xy-diagonal$ (corresponding to $\Gamma S$), $yz-diagonal$ (corresponding to $\Gamma T$), and $xyz-diagonal$ (corresponding to $\Gamma R$). 
The reasons are as follows: 
since the inverse woodpile cavity has mirror symmetry with respect to the $(y,z)$ and $(x,y)$ planes, the coupling coefficients in the $+x$ and $+z$ directions are symmetry related to those in the $-x$ and $-z$ directions, respectively, and the coefficients in the $xz-diagonal$ directions are symmetry related to each other. 
The coupling coefficients in the $+y$ and $-y$ directions are equal by reciprocity, see Appendix \ref{sec:hoppingRatesIn_yDirection}.

\begin{table*}[t]
\caption{(Nondimensional) coupling coefficients for the superlattice bands $m=3,4,5$, defined in Eq. \eqref{eq:hoppingRatesDefined} in Appendix \ref{sec:photonictightbindingmethod}. 
The calculated coupling coefficients have in addition to the real part also an imaginary part, that is at least $100\times$ smaller than the real part, that does not have physical significance, and that is not reported here. 
In addition to the 7 coefficients $\kappa$ per band (see text), we also provide the $\beta$, defined in Eq. \eqref{eq:betaCoefficient} in Appendix \ref{sec:photonictightbindingmethod}.}
\label{table:hoppingRates}
\begin{tabular}{ |p{3.0cm}|p{3.0cm}|p{3.0cm}|p{3.0cm}|  }
\hline
Coupling coefficients& $m=3$ & $m=4$ & $m=5$ \\
\hline
$\beta$         & $-1.8\cdot10^{-10}$ & $+3.3\cdot10^{-10}$ & $-1.8\cdot10^{-10}$ \\
$\kappa_{x}$                     & $+5.2\cdot10^{-4}$    & $-1.0\cdot10^{-3}$  & $+5.4\cdot10^{-4}$ \\
$\kappa_{y}$                     & $+4.7\cdot10^{-5}$    & $+4.3\cdot10^{-5}$   & $+3.0\cdot10^{-5}$ \\
$\kappa_{z}$                     & $+5.1\cdot10^{-4}$    & $-1.1\cdot10^{-3}$  & $+5.3\cdot10^{-4}$ \\
$\kappa_{xz}$ ($//~\Gamma U$)  & $-4.2\cdot10^{-4}$   & $+6.6\cdot10^{-5}$   & $-1.5\cdot10^{-4}$ \\
$\kappa_{xy}$ ($//~\Gamma S$)  & $-1.4\cdot10^{-6}$   & $+9.7\cdot10^{-6}$   & $-4.6\cdot10^{-5}$ \\
$\kappa_{yz}$ ($//~\Gamma T$)  & $-1.6\cdot10^{-6}$   & $+1.0\cdot10^{-5}$   & $-4.6\cdot10^{-5}$ \\
$\kappa_{xyz}$ ($//~\Gamma R$) & $+7.4\cdot10^{-6}$    & $+2.4\cdot10^{-6}$   & $-2.2\cdot10^{-5}$ \\
\hline
\end{tabular}
\end{table*}
\begin{figure}[h!]
\includegraphics[width=1.00\columnwidth]{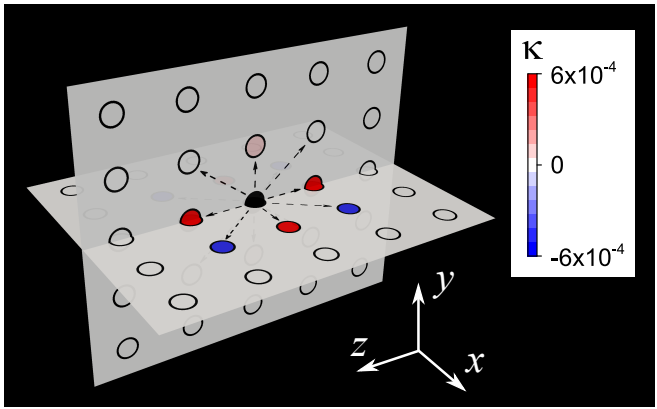}
\caption{
Coupling coefficients of light from a central cavity (black circle) to neighboring cavities (other circles) for the $m = 3$ coupled-cavity band, indicated with arrows. 
Nonzero coefficients only occur in the $(x,y)$ and $(x,z)$ planes.
Blue and red indicate negative and positive coupling coefficients, respectively, as shown by the color bar. 
The $x, y, z$ coordinate system is shown. 
This figure has been made using ParaView~\cite{Ayachit2015Book}.
}
\label{fig:hoppingRatesOfLightThirdBand}
\end{figure}

The coupling coefficients are given in Table \ref{table:hoppingRates}. 
For the $m = 3$ band the coupling coefficients of light are overlaid on the cavity superlattice structure in Figure~\ref{fig:hoppingRatesOfLightThirdBand}. 
In the $x$ and $z$ directions, the coupling coefficients are relatively large and positive, in the $xz-diagonal$ directions the coupling coefficients are large and negative, in the $y$ direction the coupling coefficient is about $10 \times$ smaller, and in all other directions the coupling coefficients are vanishingly small (typically $100\times$ less). 

Remarkably, the simultaneous occurrence of large coupling coefficients with near-vanishing dispersion bandwidths means that 1D CROW-like arguments \textit{do not hold} for 3D cavity superlattices. 
In other words, the bandwidth in a particular crystal direction for a 3D cavity superlattice is not necessarily proportional to the coupling coefficient in the same direction in real space. 
The small difference between the $x$ and $z$ coefficients confirms that the $x$ and $z$ directions are not symmetry related for the inverse-woodpile cavity, as opposed to the perfect inverse-woodpile structure.\cite{Devashish2017PRB} 

According to an 1D CROW-like argument, the large coupling coefficient in the $xz-diagonal$ directions agrees with the observation of a large dispersion bandwidth in the diagonal $U$ direction, see Fig.~\ref{fig:dispersionBandwidthThird}. 
However, the negative sign disagrees with the fact that at the $U$ point the band frequency is lower than at the $\Gamma$ point, since the reverse is true for an 1D CROW, see Ref.~\cite{Yariv1999OL}. 

The $\kappa_y$ coupling coefficient in the $y$ direction is small, in agreement with the band frequencies that are almost the same at $\Gamma$ and $Y$. 
It is remarkable that the $\kappa_y$ coupling coefficient is $10 \times$ smaller than the coefficients for the $x$ or $z$ directions, while the nearest neighbor distance is only a $(\sqrt2)$ greater than in the $x$ or $z$ directions, which would correspond to only a $(\text{exp}(\sqrt2) \approx 4 \times)$ smaller coefficient for cavities coupled by evanescent Bloch modes. 
We surmise that the quadrupolar field pattern of each cavity in the $xz$ plane couples poorly to a neighboring cavity in a neighboring $xz$ plane. 
Therefore, light mostly hops in 2D $(x,z)$-layers, which is analogous to 2D electron transport in graphite or graphene layers.\cite{Castroneto2009RMP,Jacqmin2014PRL} 
Since the light propagates very unusually by hopping only in a few discrete directions, we propose the name "Cartesian light" for the propagation of light in a 3D cavity superlattice. 



\begin{figure}[h]
\includegraphics[width=1.00\columnwidth]{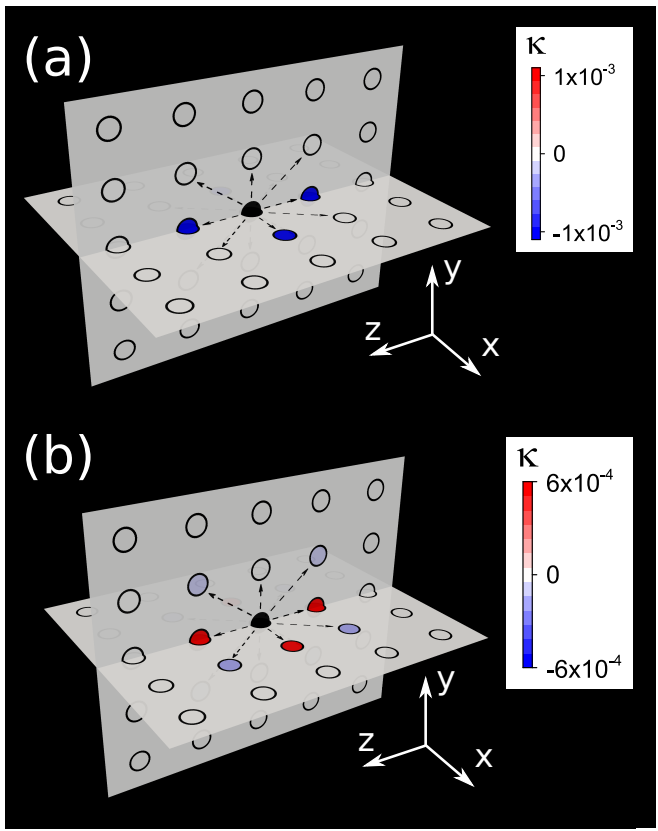}
\caption{
(a) Coupling coefficients of light from a central cavity (black circle) to neighboring cavities (other circles) for the $m = 4$ coupled-cavity band, indicated with arrows. 
Nonzero coefficients only occur in the $(x,y)$ and $(x,z)$ planes. 
Blue and red indicate negative/positive coupling coefficients, for which bonding/antibonding resonances of the two coupled cavities are energetically favorable.
The $x, y, z$ coordinate system is shown.
(b) Coupling coefficients of light from a central cavity (black circle) to neighboring cavities (other circles) for the $m = 5$ coupled-cavity band.
}
\label{fig:hoppingRatesOfLightFourthFifthBand}
\end{figure}

The coupling coefficients of light for the $m = 4,5$ bands are shown in Fig.~\ref{fig:hoppingRatesOfLightFourthFifthBand}. 
For the $m = 4$ band, the nonzero coefficients are $\kappa_x$, and $\kappa_z$. 
The coupling coefficients to all other neighboring cavities vanish, including the coefficient $\kappa_{y}$ in the $y$ direction. 
In the hopping of the $m = 4$ band we find the ultimate Cartesian light: light hops only in the $x$-$z$ directions.
For the $m = 5$ band, the nonzero coefficients are $\kappa_x$, $\kappa_y$, $\kappa_z$, $\kappa_{xz}$, $\kappa_{xy}$, $\kappa_{yz}$, and $\kappa_{xyz}$, that is, nonzero coupling coefficients to all neighboring cavities. 
In the $(x,z)$ plane, there is a mix of positive $x$ and $z$ coupling coefficients, and negative $xz-diagonal$ coupling coefficients, which is similar to the $m = 3$ band. 

We have not analyzed the $m = 1,2$ bands of coupled-cavity modes, since the band structures do not converge monotonically with increasing spatial resolution, as is elaborated in Appendix~\ref{sec:Calcs_Convrg}. 

\subsection{Propagation in 3D on the superlattice}
\label{sec:propagation-real-space}

We now discuss the propagation in real space, and why the dispersion bandwidth of the $m = 3$ band is much larger in the $U$ direction than in the other directions in the $(k_X, k_Z)$ plane in reciprocal space. 
We first discuss the dispersion bandwidth in the $xz-diagonal$ directions that are symmetry related to each other, and that correspond to the $\Gamma U$ direction in reciprocal space. 

\begin{figure}[tbp]
    \centering
    \includegraphics[width=0.5\textwidth]{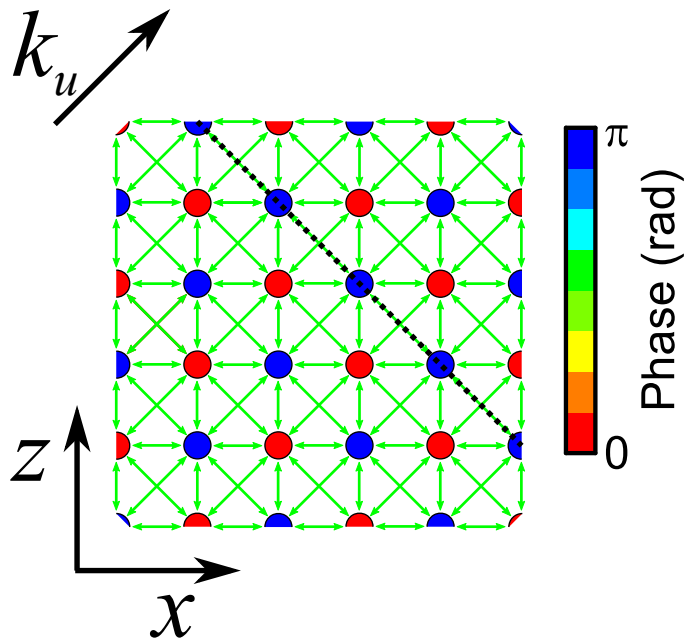}
    \caption{Relative phase of the resonating cavities for a coupled-cavity mode with a wave vector at the $U$ high-symmetry point. 
    As seen from a given cavity, the wave vector $\mathbf{k}$ is directed towards an $xz$ diagonally neighboring cavity.
    The black dashed line indicates a wavefront of the Bloch wave.
    The blue and red cavities resonate out-of-phase with each other.
    The couplings are indicated with arrows, which are green if the corresponding cavities resonate with an energetically favorable relative phase, as is the case for all couplings. 
    }
\label{fig:schematicUPointFavorable}
\end{figure}

Let us consider the relative phase of the resonating cavities for a wave vector at the $U$-point ($\mathbf{k} = \mathbf{k}_{BZ} = U$), as shown in Figure~\ref{fig:schematicUPointFavorable}. 
Since this coupled resonance eigenmode is a Bloch wave of the 3D superlattice (with a phase front as indicated in Fig.~\ref{fig:schematicUPointFavorable}) it is clear that this collective oscillation differs fundamentally from waveguiding behavior in a 1D CROW; in other words, the superlattice Bloch modes differ from the ones in a CROW. 

In Figure~\ref{fig:schematicUPointFavorable}, neighboring cavities in the $xz-diagonal$ direction resonate in-phase with each other. 
Neighboring cavities in the $x$, or $z$-direction resonate out-of-phase with each other.
Hence, there is a checkerboard pattern of two sublattices of cavities that resonate out-of-phase with each other. 
The in-phase resonance of neighboring cavities in the $xz$-diagonal direction is energetically favorable for the negative coupling coefficient in the $xz$-diagonal directions, as can be understood from Eq. \eqref{eq:hoppingRatesDefined} in Appendix \ref{sec:photonictightbindingmethod}. 
The out-of-phase resonance of neighboring cavities in the $x$, or $z$-direction is energetically favorable for the positive coupling coefficient in the $x$, or $z$-direction.
Hence, the checkerboard pattern of out-of-phase resonating cavities is energetically favorable for all coupling coefficients in the ($x$, $z$)-plane, and the band frequency is relatively low at the $U$ high-symmetry point. 

\begin{figure}[tbp]
    \centering
    \includegraphics[width=0.5\textwidth]{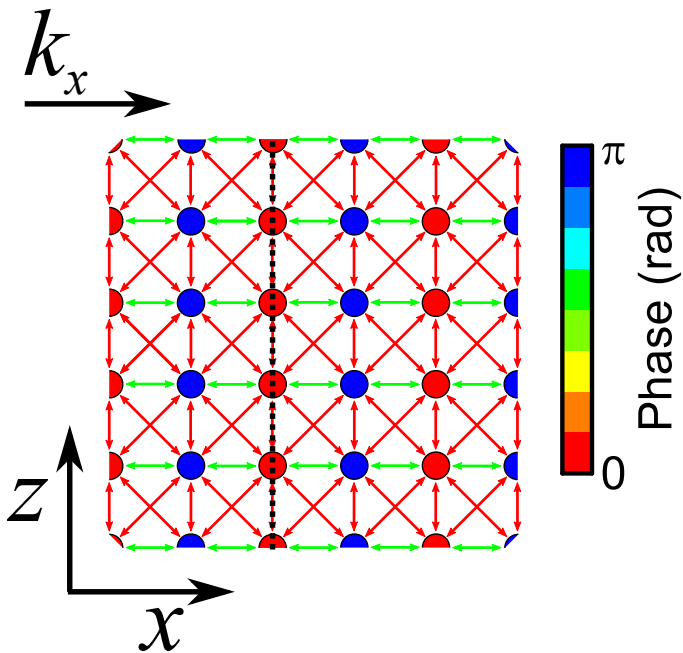}
\caption{ 
Relative phase of the resonating cavities for a coupled-cavity mode with a wave vector at the $X$ high-symmetry point. 
As seen from a given cavity, the wave vector $\mathbf{k}$ is directed towards a nearest-neighboring cavity. 
The black dashed line indicates a wavefront of the Bloch wave.
The blue and red cavities resonate out-of-phase with each other.
The couplings are indicated with arrows, green arrows indicate an energetically favorable relative phase, and red arrows indicate an unfavorable phase. 
}
\label{fig:schematicXPointFavorable}
\end{figure}

We now consider the superlattice Bloch mode at the $\mathbf{k} = \Gamma$ point (the origin in reciprocal space). 
All cavities resonate in-phase with each other, which is energetically favorable for the negative coupling coefficient in the $xz$-diagonal direction. 
However, it is not energetically favorable for the positive coupling coefficients in the $x$, or $z$-direction.
Hence, the band frequency is relatively high at the $\Gamma$ point, and higher than at the $U$-point, in agreement with the observation of large dispersion bandwidth in Figure~\ref{fig:bandStructureCrystalOfCavities}(b). 


We now discuss the dispersion bandwidth in any direction in the $(k_X, k_Z)$ plane other than the $xz$-diagonal direction. 
For example, the relative phase of the resonating cavities at the $X$ high-symmetry point is shown in Fig.~\ref{fig:schematicXPointFavorable}.
Only the checkerboard pattern of out-of-phase resonating cavities at the $U$ high-symmetry point is energetically favorable for all coupling coefficients in the ($x$, $z$) plane.
For any other point in the Brillouin zone, the relative phase of the resonating cavities is not energetically favorable for all coupling coefficients in the ($x$, $z$) plane. 
Hence, the dispersion bandwidth is small for directions other than the $xz$-diagonal direction, in agreement with the superlattice band structures in Figure~\ref{fig:bandStructureCrystalOfCavities}(b). 

\section{Discussion}\label{sec:discussion}

\subsection{Differences with other forms of light transport} 

Cartesian light propagation on a 3D superlattice of cavities differs fundamentally from other known modes of propagation in periodic nanophotonic systems, notably from the conventional 3D spatially-extended Bloch wave propagation in crystals, from light tunneling through a band gap, from coupled-resonator optical waveguiding, and also from light diffusing at the edge of a gap. 

(1) A characteristic feature of the 3D superlattice Bloch modes is that they are constructed from modes where the light field is hopping from lattice site to lattice site. 
In other words, the field pattern has its maxima on the lattice sites (i.e. the cavities) and decays exponentially in between lattice sites, since isolated cavities are tuned inside the photonic band gap where the wave vector is complex. 
In contrast, the Bloch modes outside the photonic band gap are constructed from purely real modes of propagation; there is no reason for field maxima to be located on preferred positions in the crystal. 

(2) A second characteristic feature of the 3D superlattice Bloch modes is that they are genuine modes of propagation centred within the 3D photonic band gap. 
In this sense, they are distinguished from the modes in photonic crystals with finite support that were recently described in Ref.~\cite{Hasan2018PRL}. 
In that study, it was found that the finite extent of a photonic band gap crystal leads to the filling of the density of states (DOS) in the band gap by states that are centred outside the band gap, while extending into the band gap due to their substantial band width. 

(3) The 3D Cartesian superlattice propagation differs fundamentally from the propagation in lower-dimensional 1D (a CROW) and 2D arrays of cavities. 
Firstly, in section~\ref{sec:propagation-real-space} we have already discussed that the superlattice modes differ fundamentally from those of a CROW. 
In other words, a 3D superlattice does not seem to be a "3D CROW". 
Secondly, if we perturb the frequency of one of the cavities in a superlattice, a bound state appears instantaneously in 1D and 2D, whereas a threshold frequency difference is required in 3D, see Ref.~\cite{Economou2006Book}. 

(4) The propagation of light in a 3D cavity superlattice in a photonic band gap differs fundamentally from directional diffusion that was identified for 3D photonic band gap crystals with a certain degree of disorder~\cite{Koenderink2003PhysRevLett}. 
In the latter case, the modes of propagation are not waves but diffusive. 
Moreover, the typical frequencies are at the edge of the band gap, hence outside the gap, as opposed to the cavity superlattice modes that reside within the band gap, see Fig.~\ref{fig:bandStructureCrystalOfCavities}.

\subsection{Crystal structures of the cavity superlattice } 
If the magnification factors of the superlattice's lattice parameters, compared to the underlying crystal structure's lattice parameters, fulfill $M_i \neq (M_j, M_k)$ ($i,j,k = x,y,z$), the cavity superlattice is not cubic anymore - in contrast to the underlying inverse woodpile structure - but has become tetragonal. 
In the most general case with $M_x \neq M_y \neq M_z$ the superlattice has different cavity spacings in each direction $(x,y,z)$; the superlattice has then become orthorhombic. 
Given that cavities in an inverse woodpile structure are necessarily located along the smaller-pore line defects, we currently doubt whether it is feasible to realize other 3D Bravais superlattices. 

We have seen that for the $m = 3, 4, 5$ bands, the coupling coefficient in the $y$-direction is smaller than in the $(x,z)$ plane, by typically $10 \times$. 
To make the hopping of light more 3D, it is necessary to increase the coupling coefficient in the $y$-direction compared to the coefficients in the $(x,z)$-directions, which can be achieved by a closer cavity spacing in the  $y$-direction, for example, in a $M_x M_y M_z = 3\times2\times3$ supercell in case of $M_x = M_z = 3$ (as studied here) or in general for supercells with $M_y < (M_x, M_z)$. 

Conversely, if it is desired to realize a superlattice with effectively 2D transport of light in $(x, z)$ planes, the superlattice parameter in the $y$-direction should be made greater than the ones in the $(x, z)$-directions ($M_y >> (M_x, M_z)$). 
In this situation, the 2D transport of light may hold analogies to that of charge carriers in graphite layers\cite{Castroneto2009RMP} or in high-T$_c$ superconductors. 

\subsection{Disorder} 
We have studied the dispersion and hopping for superlattices without disorder. 
Let us briefly comment on the sensitivity of the results to a small degree of disorder, since we performed calculations for several grid resolutions in Appendix~\ref{sec:Calcs_Convrg}, and a change of the grid resolution implies a slight shift in the geometry. 
On the one hand, we observed that for all five bands of coupled-cavity modes, the center frequency is highly sensitive to the grid resolution and therefore to disorder. 
This is likely the result of the lightning rod effect of the cavity mode field pattern identified in Ref.~\cite{Woldering2014PRB}, wherein the inverse woodpile cavity resonances have regions of high intensity at sharp corners in the dielectric material. 
If such sharp corners are slightly distorted, it is quite conceivable that the overlap with the field pattern changes, leading to a change in resonance frequency. 
On the other hand, we observe for bands $m = 3, 4, 5$ that the features of the dispersion bands remain the same while the grid resolution is increased. 
Therefore, we expect the coupling coefficients for the $m = 3, 4, 5$ bands to be robust to small degrees of disorder. 

A 3D cavity superlattice is the photonic analogue of the Anderson model for spins and electrons~\cite{Anderson1958PR}, albeit in the limit of zero disorder. 
The 3D cavity superlattice also corresponds to the Hubbard model without interactions~\cite{Hubbard1963ProcRSocA,Hubbard1964ProcRSocA}.
We anticipate that the present study may form the basis for further exploration of the physics of the 3D Anderson model for nanophotonic cavity superlattices that will proceed by introducing controlled degrees of disorder in the cavity resonance frequencies. 

\subsection{Outlook}

A possible application would be a scalable, coherently linked network of NV-based registers, see Ref.~\cite{Childress2013MRSBull}.





\section{Summary}
We have studied for the first time ever the propagation of light in a 3D cavity superlattice within a 3D photonic band gap.
Such a 3D cavity superlattice is the photonic analogue of the Anderson model in the limit of zero disorder. 
The light hops only in a few high-symmetry directions including the Cartesian $(x,y,z)$ directions, therefore we propose the name "\textit{Cartesian light}".
3D Cartesian hopping of light in a 3D band gap yields propagation as superlattice Bloch modes that differ fundamentally from the conventional 3D spatially-extended Bloch wave propagation in crystals, from light tunneling through a band gap, from coupled-resonator optical waveguiding, and also from light diffusing at the edge of a gap. 
The large coupling coefficients in the Cartesian directions occur simultaneously with a near vanishing dispersion bandwidth in this direction. 
This means that 1D CROW-like \textit{do not} arguments hold for 3D cavity superlattices.
The unusually small dispersion bandwidth in the Cartesian directions is a result of interplay between positive and negative coupling coefficients in the Cartesian and diagonal directions.

\section{Acknowledgments}
We thank Bill Barnes (Exeter, Twente) for coining the term "Cartesian light" and for discussions, and we thank Geert Brocks, Ad Lagendijk, Jan Kl{\"a}rs and Bart van Tiggelen for useful discussions. 
This research is supported by the 4TU federation, by the FOM/NWO programme ``Stirring of light!," the STW/NWO-Perspectief program ``Free-form scattering optics", the  ``Descartes-Huygens" prize of the French Academy of Sciences, and the MESA$^{+}$ Institute for Nanotechnology section Applied Nanophotonics (ANP).

\appendix

\section{Brillouin zone and tetragonal unit cell}
\label{sec:BrillouinZoneTetragonalUnitCell}

\begin{figure}[H]
\begin{center}
\includegraphics[scale=0.40]{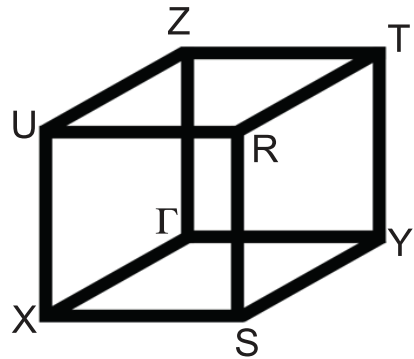}
\end{center}
\caption{
First Brillouin zone of the inverse woodpile crystal structure showing the high symmetry points $X,Y,Z,S,R,T,U$ and the origin called $\Gamma$. }
\label{fig:BrillouinZoneTetragonalUnitCell}
\end{figure}

Figure~\ref{fig:crystalOfCavities}(a) shows a tetragonal representation of the $M_x M_y M_z = 3 \times 3 \times 3$ supercell of the cubic inverse woodpile crystal structure including a pair of proximal defect nanopores that form a resonant cavity. 
In terms of the unit vectors of the conventional cubic diamond structure,\cite{Ashcroft1976Book} the tetragonal unit supercell has unit vectors $a_{x} = \frac{1}{\sqrt{2}}[M_x~0~M_x]$, $a_{y} = [0~M_y~0]$, and $a_{z} = \frac{1}{\sqrt{2}}[\bar{M_z}~0~M_z]$. 

Figure~\ref{fig:BrillouinZoneTetragonalUnitCell} shows the first Brillouin zone of the inverse woodpile cavity superlattice with characteristic high-symmetry points. 
The main axes are given by $X = [\pi/c_s^{x}, 0, 0]$, $y = [0, \pi/a_s, 0]$, and $Z = [0, 0, \pi/c_s^{z}]$. 
The $\Gamma X$ and $\Gamma Z$ directions are notable as they correspond to waves propagating along each set of nanopores, while the diagonal $\Gamma U$ direction lies in between. 
Due to the geometry of the cavity (composed of two proximal defect pores), the $\Gamma X$ and $\Gamma Z$ directions are not symmetry related, in contrast to the underlying inverse woodpile crystal structure.\cite{Devashish2017PRB}

\section{Calculations and convergence}
\label{sec:Calcs_Convrg}

We have calculated the band structure of 3D cavity superlattices using the plane-wave expansion method,\cite{Ashcroft1976Book,Joannopoulos2008Book} where we employed the well-known MIT Photonic−Bands (MPB) package~\cite{Johnson2001OptExpress}. 
We performed calculations with increasing spatial grid resolutions of $12\times17\times12$, $24\times34\times24$, $48\times68\times48$, and $96\times136\times96$ per unit cell of the underlying inverse woodpile crystal. 
Although the $12\times17\times12$ calculation is a replication of Ref.~\cite{Woldering2014PRB}, the results do not agree perfectly since the subpixel averaging has been updated in Version 1.5 of the MPB code that we use here. 
We verified that with an older version of MPB our calculations agree exactly with Ref.~\cite{Woldering2014PRB}. 

\begin{figure}[H]
\includegraphics[width=1.00\columnwidth]{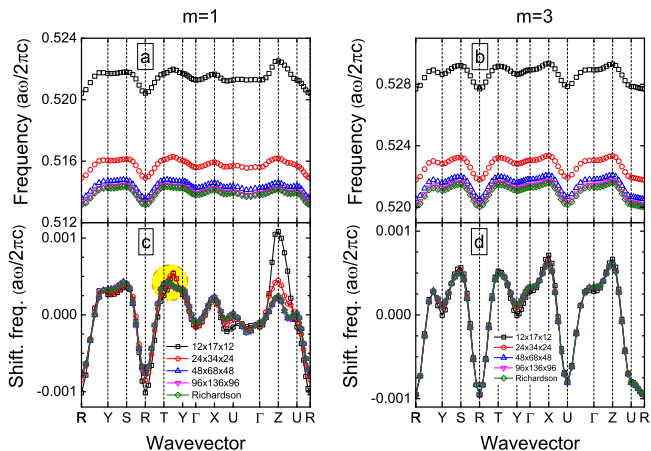}
\caption{
(a,b) Dispersion relations of the $m = 1,3$ bands of coupled-cavity modes, respectively, for grid resolutions $12\times17\times12$ (black squares), $24\times34\times24$ (red circles), $48\times68\times48$ (blue upward triangles), and $96\times136\times96$ (magenta inverted triangles). 
The Richardson extrapolated data are shown as green diamonds. 
(c,d) Overlaid dispersion relations where the center frequency of each band is subtracted. 
}
\label{fig:convergenceFirstBands1and3}
\end{figure}

To study the convergence of the defect bands with increasing spatial resolution, we show in Figure~\ref{fig:convergenceFirstBands1and3}(a,b) the $m = 1$ and $m = 3$ defect bands, respectively, for all resolutions considered. 
The general trend observed in the data - also seen with the other three defect bands - is that the average frequency of the bands decreases with increasing resolution, where the initial decrease is fast, whereas for increasing resolution the decrease slows down, and the Richardson-extrapolated (or converged) band frequencies are reached at a resolution of $96 \times 136 \times 96$. 


The initial refinement from $12\times17\times12$ to $24\times34\times24$ is remarkably large in comparison to the bandwidth of each defect band. 
We speculate that the shifts are related to the "lighting rod"-behavior of the field patterns of a cavity mode, whereby the high fields are concentrated at sharp corners~\cite{Woldering2014PRB}. 
A change in the grid resolution effectively corresponds to a change in the geometry. 
At low resolution, the field pattern likely "misses" the sharp features of the high-index material, whereas with increased resolution the field pattern increasingly fits within the high-index material, causing a decrease of the field energy - in view of the variational principle~\cite{Joannopoulos2008Book} - as observed in Figure~\ref{fig:convergenceFirstBands1and3}. 

To quantify the rate of convergence, we calculate the convergence order following Ref.~\cite{Roache1998Book}. 
We assume that the grid spacing $h$ is sufficiently refined for the error $E$ to asymptotically approach zero as 
\begin{equation}
E = \omega(h) - \omega_{\text{exact}} = C h^p,
\end{equation}
with $\omega(h)$ the frequency calculated for grid spacing $h$ at a given wave vector, $\omega_{\text{exact}}$ the exact frequency at the same wave vector, $C$ a constant. 
The convergence order $p$ is obtained from a sequence of three resolutions as\cite{Roache1998Book} 
\begin{equation}
p = ln\left(\frac{\omega(h)-\omega(\frac{h}{2})}{\omega(\frac{h}{2})-\omega(\frac{h}{4})}\right)/\sqrt{2},
\end{equation}
where we used the $12\times17\times12$, $24\times34\times24$ and $48\times68\times48$ resolution results. 
From data as shown in Figure~\ref{fig:convergenceFirstBands1and3}(a,b), we obtain 
$p = 1.85, 1.96, 2.24, 2.16, 2.10$ for $m = 1, 2, 3, 4, 5$, respectively, in close agreement with the convergence order $p = 2$ for the plane-wave expansion method.~\cite{Johnson2001OptExpress} 
We therefore conclude that the band frequencies are converging accurately, as expected.

Moreover, the convergence of the band frequencies allows us to use Richardson extrapolation to obtain the band frequency $\omega_{h=0}$ in the limit of infinite grid resolution ($h = 0$).\cite{Roache1998Book,Richardson1927PhilosTransRoyalSocA} 
The frequency $\omega_{h=0}$ is estimated as 
\begin{equation}
\omega_{h=0} = \omega\Big(\frac{h}{4}\Big) + \frac{\omega(\frac{h}{4}) - \omega(\frac{h}{2})}{2^p-1}. 
\label{eq:estd_infinite_resolution}
\end{equation}
The Richardson-extrapolated frequencies are shown in Fig.~\ref{fig:convergenceFirstBands1and3}(a,b) as a function of wave vector for the $m = 1$ and $m = 3$ defect bands, respectively. 
The Richardson-extrapolated frequencies are slightly below the frequencies for the finest grid resolution ($96 \times 136 \times 96$) as expected in case of convergence~\cite{Roache1998Book}. 

Nevertheless, unexpected features were found in the dispersion relations as a function of spatial grid resolution. 
Figures~\ref{fig:convergenceFirstBands1and3}(c,d) show the dispersion relations of the $m = 1$ and $m = 3$ defect bands that are overlaid for all grid resolutions by subtracting the average band frequencies from the data in Figures~\ref{fig:convergenceFirstBands1and3}(a,b). 
For the $m = 1$ defect band, we observe that for resolutions $12\times17\times12$ and $24\times34\times24$, the band between the $T$ and $Y$ high-symmetry points has a maximum at the $4^{\text{th}}$ symbol from the left, midway in between $T$ and $Y$.
For $48\times68\times48$ and $96\times136\times96$, the maximum has moved to the $1^{\text{st}}$ symbol from the left, or one sixth of the way from $T$ to $Y$. 
Thus, there is a qualitative change in the dispersion relations as the grid is refined. 
Similar observations were made on the $m = 2$ defect band. 
Therefore, we do not trust the $m = 1,2$ defect bands sufficiently to derive coupling coefficients. 
In contrast, for the $m = 3$ defect band, the maximum in frequency is always at the $T$ point, thus the shape is preserved, and the band readily converges. 
Similar observations as for $m = 3$ were made on the $m = 4,5$ defect bands. 
Therefore, we trust the convergence of the $m = 3,4,5$ bands sufficiently to warrant the extraction of coupling coefficients.


\section{Photonic tight-binding method}
\label{sec:photonictightbindingmethod}
We employ the well-known tight-binding method to model the dispersion of the defect bands~\cite{Ashcroft1976Book,Yariv1999OL}.
In developing the tight-binding approximation, we assume that in the vicinity of each lattice point the full periodic superlattice dielectric function, $\epsilon$, can be approximated by the dielectric function, $\epsilon_{\Omega}$, of a single cavity located at the lattice point.
This assumption is valid for the cavity superlattice, since the superlattice dielectric function differs only from $\epsilon_{\Omega}$ at the defect pores in the superlattice that do not form part of the cavity, all of which are at least a lattice constant away.
We also assume that the modes of the cavity are well localized; i.e., if $\mathbf{E}_{\Omega}^m$ is a mode of a cavity at the origin,
\begin{equation} \label{eq:cavityMaxwellEquations}
\nabla \times \nabla \times \mathbf{E}_{\Omega}^m = \Big( \frac{\Omega_m}{c} \Big)^2 \epsilon_{\Omega}(\mathbf{r})\mathbf{E}_{\Omega}^m,
\end{equation}
with $\Omega_m$ the resonance frequency of the single cavity and $c$ the speed of light, then we require that $\mathbf{E}_{\Omega}^m(\mathbf{r})$ be very small when $r$ exceeds a distance of the order of the lattice constant, which we shall refer to as the "range" of $\mathbf{E}_{\Omega}^m$.
Let us briefly verify this assumption: for the defect bands, the mode volume is about $V \approx \lambda^3$, with $\lambda$ the free-space wavelength~\cite{Woldering2014PRB}. 
Hence, the electric field attenuates over a range of about $\lambda / 2$ in a given direction. 
We model the cavity mode as an evanescent plane wave that attenuates by a factor of $1/e$ over this distance, which yields an imaginary part of the wave vector of $k'' = 2/\lambda$.
The nearest neighbor distances are equal to the lattice parameters of the superlattice $(c_s^x, a_s, c_s^z)$ that are multiples of the inverse woodpile lattice parameters $(c_s^x, a_s, c_s^z) = (3c, 3a, 3c)$ (with $c=a/\sqrt{2}$), and the reduced frequency $\omega'=a/\lambda$ of a cavity resonance is equal to $\omega'=0.52 ... 0.54$ (for $m = 3, 4, 5$). 
For the directions $x$ or $z$, this yields a product of the imaginary part of the wave vector $k''$ and the nearest neighbor distance $\Delta r$ of 
\begin{equation}
k'' \Delta r = \frac{2}{\lambda} . c_s^{x,z} = \frac{2}{{\frac{a}{\omega'}}} . 3 c = 
6 \frac{\omega'}{\sqrt{2}} \approx 2.2.
\end{equation}
Thus the electric field intensity from one cavity has attenuated to as little as $\mathrm{exp}(-2.2) \approx 0.11$ at a nearest neighboring cavity in the directions $x$ or $z$, and for the $y$-direction the decay is even greater, thereby readily fulfilling the requirements of the tight-binding method.

In the tight-binding method, we write the superlattice dielectric function $\epsilon$ as
\begin{equation} \label{eq:crystalMaxwellEquations}
\epsilon = \epsilon_{\Omega} + \Delta \epsilon(\mathbf{r}),
\end{equation}
where $\Delta \epsilon(\mathbf{r})$ contains all corrections to the cavity dielectric function required to produce the full periodic dielectric function of the superlattice.
Since the product $\Delta \epsilon(\mathbf{r}) \mathbf{E}_{\Omega}^m(\mathbf{r})$, though nonzero, is exceedingly small, we might expect the solution to the full superlattice Maxwell equations to be quite close to the cavity wave function $\mathbf{E}_{\Omega}^m(\mathbf{r})$ or to wave functions with which $\mathbf{E}_{\Omega}^m(\mathbf{r})$ is degenerate.
Based on this expectation, one seeks an $\mathbf{E}(\mathbf{r})$ that can be expanded in a relatively small number of localized cavity wave functions:
\begin{equation} \label{eq:AshcroftMermin10.6and10.7combined}
\mathbf{E}(\mathbf{r}) = \sum_{\mathbf{R}} e^{i\mathbf{k}\cdot \mathbf{R}} \sum_m b_m \mathbf{E}_{\Omega}^m(\mathbf{r} - \mathbf{R}).
\end{equation}
The wave vector $\mathbf{k}$ describes the relative phase of the resonating cavities and ensures that $\mathbf{k}$ satisfies the Bloch theorem for the Brillouin zone of the superlattice of cavities.

If we multiply the superlattice Maxwell equations
\begin{equation}
\nabla \times \nabla \times \mathbf{E}(\mathbf{r}) = \Big( \frac{\omega}{c}\Big)^2 [\epsilon_{\Omega}(\mathbf{r}) + \Delta \epsilon(\mathbf{r})] \mathbf{E}(\mathbf{r})
\end{equation}
by the cavity wave function $\mathbf{E}_{\Omega}^{n*}(\mathbf{r})$ where the star represents the complex conjugate, integrate over all $\mathbf{r}$, and use the fact that
\begin{align}
&\int_{-\infty}^{\infty} \mathbf{E}_{\Omega}^{n*}(\mathbf{r})\cdot \nabla \times \nabla \times \mathbf{E}(\mathbf{r}) \text{d}\mathbf{r} \nonumber \\
&= \int_{-\infty}^{\infty} [ \nabla \times \nabla \times \mathbf{E}_{\Omega}^n (\mathbf{r})]^* \cdot \mathbf{E}(\mathbf{r})\text{d}\mathbf{r} \nonumber \\
&= \Big( \frac{\Omega_n}{c} \Big)^2 \int \epsilon_{\Omega}(\mathbf{r}) \mathbf{E}_{\Omega}^{n*}(\mathbf{r}) \cdot \mathbf{E} (\mathbf{r}) \text{d} \mathbf{r},
\end{align}
we find that
\begin{align} \label{eq:AshcroftMermin10.10}
\bigg[ \Big( \frac{\Omega_n}{c} \Big)^2 - \Big( \frac{\omega(\mathbf{k})}{c}\Big)^2 \bigg] \int_{-\infty}^{\infty} \epsilon_{\Omega}(\mathbf{r}) \mathbf{E}_{\Omega}^{n*}(\mathbf{r})\cdot \mathbf{E}(\mathbf{r})\text{d} \mathbf{r} \nonumber \\
= \Big( \frac{\omega(\mathbf{k})}{c}\Big)^2 \int_{-\infty}^{\infty} \mathbf{E}_{\Omega}^{n*}(\mathbf{r})\cdot \Delta \epsilon(\mathbf{r}) \mathbf{E}(\mathbf{r})\text{d}\mathbf{r}.
\end{align}
Inserting Eq.~\eqref{eq:AshcroftMermin10.6and10.7combined} into Eq.~\eqref{eq:AshcroftMermin10.10} and using the orthonormality of the cavity wave functions,
\begin{equation}
\int_{-\infty}^{\infty} \epsilon_{\Omega}(\mathbf{r})\mathbf{E}_{\Omega}^{n*} (\mathbf{r}) \cdot \mathbf{E}_{\Omega}^m(\mathbf{r}) \text{d} \mathbf{r} = \delta_{mn},
\end{equation}
we arrive at an eigenvalue equation that determines the coefficients $b_m(\mathbf{k})$ and the Bloch frequencies $( \omega(\mathbf{k})/c)^2$:
\begin{align} \label{eq:AshcroftMermin10.12}
&\bigg[ \Big( \frac{\Omega_n}{c}\Big)^2 - \Big( \frac{\omega(\mathbf{k})}{c}\Big)^2\bigg]b_n \nonumber \\
&= -\bigg[ \Big( \frac{\Omega_n}{c}\Big)^2 - \Big( \frac{\omega(\mathbf{k})}{c}\Big)^2\bigg] \sum_m\nonumber\\
&\bigg( \sum_{\mathbf{R} \neq \mathbf{0}} \int_{-\infty}^{\infty} \mathbf{E}_{\Omega}^{n*}(\mathbf{r}) \cdot \epsilon_{\Omega}(\mathbf{r}) \mathbf{E}_{\Omega}^m(\mathbf{r} - \mathbf{R})e^{i \mathbf{k}\cdot \mathbf{R}} \text{d} \mathbf{r} \bigg) b_m \nonumber\\
&+ \Big( \frac{\omega(\mathbf{k})}{c}\Big)^2 \sum_m \bigg( \int_{-\infty}^{\infty} \mathbf{E}_{\Omega}^{n*}(\mathbf{r}) \cdot \Delta \epsilon(\mathbf{r}) \mathbf{E}_{\Omega}^m(\mathbf{r}) \text{d} \mathbf{r}  \bigg) b_m \nonumber\\
&+ \Big( \frac{\omega(\mathbf{k})}{c}\Big)^2 \sum_m \nonumber\\
&\bigg( \sum_{\mathbf{R} \neq \mathbf{0}} \int_{-\infty}^{\infty} \mathbf{E}_{\Omega}^{n*}(\mathbf{r}) \cdot \Delta \epsilon(\mathbf{r}) \mathbf{E}_{\Omega}^m(\mathbf{r} - \mathbf{R})e^{i \mathbf{k} \cdot \mathbf{R}} \text{d} \mathbf{r}\bigg) b_m.
\end{align}

The first term on the right of Eq. \eqref{eq:AshcroftMermin10.12} contains integrals of the form
\begin{equation} \label{eq:AshcroftMermin10.13}
\int_{-\infty}^{\infty} \text{d} \mathbf{r} \mathbf{E}_{\Omega}^{n*}(\mathbf{r}) \cdot \epsilon_{\Omega}(\mathbf{r}) \mathbf{E}_{\Omega}^m(\mathbf{r} - \mathbf{R}).
\end{equation}
We interpret our assumption of well-localized cavity modes to mean that Eq.~\eqref{eq:AshcroftMermin10.13} is small compared to unity.
We assume that the integrals in the third term on the right of Eq.~\eqref{eq:AshcroftMermin10.12} are small, since they also contain the product of two cavity wave functions centered at different sites.
Finally, we assume that the second term on the right of Eq.~\eqref{eq:AshcroftMermin10.12} is small because we expect the cavity wave functions to become small at distances large enough for the periodic dielectric function to deviate appreciably from the cavity one.

Consequently, the right-hand side of \eqref{eq:AshcroftMermin10.13} (and therefore $[ ( \Omega_n/c )^2 - ( \omega(\mathbf{k})/c)^2 ]b_n$) is always small.
This is possible if $( \Omega_n/c)^2 - ( \omega(\mathbf{k})/c)^2$ is small whenever $b_n$ is not (and vice versa).
Thus $( \omega(\mathbf{k})/c)^2$ must be close to a cavity mode, say $( \Omega_0/c)^2$, and all the $b_n$ except those going with that mode and modes degenerate with (or close to) it in frequency must be small:
\begin{align} \label{eq:AshcroftMermin10.14}
&\Big( \frac{\omega}{c} \Big)^2 (\mathbf{k}) \approx \Big( \frac{\Omega_0}{c}\Big)^2, \enskip b_n \approx 0 \enskip \text{unless} \nonumber \\
&\Big(\frac{\Omega_n}{c} \Big)^2 \approx \Big( \frac{\Omega_0}{c}\Big)^2.
\end{align}

We can exploit Eq.~\eqref{eq:AshcroftMermin10.14} to estimate the right-hand side of \eqref{eq:AshcroftMermin10.12} by letting the sum over $m$ run only through those modes with frequencies either degenerate with or very close to $( \Omega_0/c)^2$.
If the cavity mode $0$ is nondegenerate, i.e. an $s$-orbital like mode, then in this approximation \eqref{eq:AshcroftMermin10.12} reduces to a single equation giving an explicit expression for the frequency of the band arising from this $s$-orbital like mode (generally referred to as an "$s$-orbital like band")
\begin{equation} \label{eq:AshcroftMermin10.15}
\Big( \frac{\omega(\mathbf{k})}{c}\Big)^2 = \Big( \frac{\Omega_s}{c}\Big)^2 - \frac{\beta_k + \sum \gamma_k (\mathbf{R})e^{i\mathbf{k}\cdot \mathbf{R}}}{1+\sum \alpha(\mathbf{R})e^{i\mathbf{k}\cdot \mathbf{R}}},
\end{equation}
where $( \Omega_s/c)^2$ is the frequency of the cavity $s$-orbital like mode, and
\begin{align} 
\beta_k &\equiv \Big( \frac{\omega(\mathbf{k})}{c}\Big)^2 \int_{-\infty}^{\infty} \text{d}\mathbf{r} \mathbf{E}_{\Omega}^*(\mathbf{r}) \cdot \Delta \epsilon(\mathbf{r}) \mathbf{E}_{\Omega}(\mathbf{r}),\\
\alpha(\mathbf{R}) &\equiv \int_{-\infty}^{\infty} \text{d}\mathbf{r} \mathbf{E}_{\Omega}^*(\mathbf{r}) \cdot \epsilon_{\Omega}(\mathbf{r}) \mathbf{E}_{\Omega}(\mathbf{r} - \mathbf{R}),\\
\gamma_k(\mathbf{R}) &\equiv \Big( \frac{\omega(\mathbf{k})}{c}\Big)^2 \int_{-\infty}^{\infty} \text{d} \mathbf{r} \mathbf{E}_{\Omega}(\mathbf{r}) \cdot \Delta \epsilon(\mathbf{r}) \mathbf{E}_{\Omega}(\mathbf{r} - \mathbf{R}).
\end{align}
Here the $\gamma_k(\mathbf{R})$ coefficient is the optical analogue of the hopping integral known from the electronic case.
In comparison to the tight-binding dispersion relation for Schr\"{o}dinger waves (see Ref.~\cite{Ashcroft1976Book} equation (10.15)), we note a remarkable difference, namely that the integral here is multiplied with the frequency.
We infer that a physical reason for this difference is that electronic potentials can be arbitrarily high or low compared to kinetic energy, whereas the optical analogue "potential for light" is always below the photon energy level, see for illustration the 2nd Figure of Ref.~\cite{John1991PT}.

We neglect the terms in $\alpha$ in the denominator of Eq.~\eqref{eq:AshcroftMermin10.15}, since they give only small corrections to the numerator. 
Another simplification consists of assuming that only nearest-neighbor separations give appreciable overlap integrals. 
Combining these assumptions, we simplify Eq.~\eqref{eq:AshcroftMermin10.15} to 
\begin{equation}
\Big(\frac{\omega(\mathbf{k})}{c}\Big)^2 = \Big( \frac{\Omega_s}{c}\Big)^2 - \beta_k - \sum_{\text{n.n.}} \gamma_k(\mathbf{R})e^{i\mathbf{k}\cdot \mathbf{R}},
\end{equation}
where the sum runs only over those $\mathbf{R}$ in the Bravais lattice that connect the origin to its nearest neighbors.
An explicit expression for the frequency is
\begin{equation}
\Big( \frac{\omega(\mathbf{k})}{c}\Big)^2 = \Big( \frac{\Omega_s}{c} \Big)^2 \frac{1}{\sqrt{1+\beta + \sum_{\text{n.n.}} \kappa(\mathbf{R})e^{i\mathbf{k}\cdot \mathbf{R}}}},
\end{equation}
with
\begin{align}
\beta &\equiv \int_{-\infty}^{\infty} \text{d} \mathbf{r} \mathbf{E}_{\Omega}^*(\mathbf{r}) \cdot \Delta \epsilon(\mathbf{r}) \mathbf{E}_{\Omega}(\mathbf{r}),\label{eq:betaCoefficient}\\
\kappa(\mathbf{R}) &\equiv \int_{-\infty}^{\infty} \text{d} \mathbf{r} \mathbf{E}_{\Omega}^*(\mathbf{r})\cdot \Delta \epsilon(\mathbf{r}) \mathbf{E}_{\Omega}(\mathbf{r} - \mathbf{R}).\label{eq:hoppingRatesDefined}
\end{align}
The $\kappa(\mathbf{R})$ coefficients are referred to as the coupling coefficients of light in our paper, and are nondimensional.
Since we verified that $\mathbf{E}_{\Omega}(\mathbf{r})$ is small at distances exceeding the lattice parameter of the superlattice, the coefficient in Eq. \eqref{eq:betaCoefficient} is negligible.
To obtain an expression for the frequency rather than the frequency squared, we take the square root, and we apply the first-order Taylor approximation $(1+x)^{-\frac{1}{2}}=1-x/2+O(|x|^2)$, which leads to the dispersion relation
\begin{equation} \label{eq:tightBindingDispersionRelation}
\frac{\omega(\mathbf{k})}{c} = \frac{\Omega_s}{c}\bigg( 1 - \frac{\beta}{2} - \frac{1}{2}\sum_{\text{n.n.}} \kappa(\mathbf{R}) e^{i\mathbf{k}\cdot \mathbf{R}}\bigg)
\end{equation}

The coupling coefficients are extracted from the dispersion by means of a least-squares fit of the right-hand side of Eq.~\eqref{eq:tightBindingDispersionRelation} to the dispersion throughout the whole first Brillouin zone. 
We represent the first Brillouin zone by a grid of $12\times12\times12$ cubes.
We fit over the $\mathbf{k}$ vectors in the middle of the cubes.
The results of the tight-binding calculations are discussed in Section \ref{sec:results}.
We note that for all nonzero coupling coefficients, the imaginary part is of the order of at most $10^{-6}$, which is $100\times$ to $1000\times$ less than the real part (of the order of $ \sim 10^{-4}$ for $m=3,5$ and $\sim 10^{-3}$ for $m=4$). 
Therefore, we are confident that the coupling coefficients are physically significant. 
As expected, $\beta$ is negligibly small, with a real part of the order of $\sim 10^{-10}$. 

\section{Coupling coefficients in the $y$-direction}
\label{sec:hoppingRatesIn_yDirection}
The coupling coefficients in the $+y$ and $-y$-directions are equal by reciprocity which we derive in the following.
We start with the definition of the coupling coefficient
\begin{align}
\kappa(\mathbf{R}) &= \int \text{d}\mathbf{r} \mathbf{E}_{\Omega}^*(\mathbf{r})\cdot \Delta \epsilon(\mathbf{r}) \mathbf{E}_{\Omega}(\mathbf{r} - \mathbf{R}).
\intertext{By definition}
&= \int \text{d} \mathbf{r} \mathbf{E}_{\Omega}^*(\mathbf{r})\cdot [\epsilon(\mathbf{r}) - \epsilon_{\Omega}(\mathbf{r})] \mathbf{E}_{\Omega}(\mathbf{r} - \mathbf{R}).
\intertext{Translating the coordinate system}
&= \int \text{d} \mathbf{r} \mathbf{E}_{\Omega}^*(\mathbf{r} + \mathbf{R}) \cdot [\epsilon(\mathbf{r} + \mathbf{R}) - \epsilon_{\Omega}(\mathbf{r} + \mathbf{R})] \mathbf{E}_{\Omega}(\mathbf{r}).
\intertext{By Maxwell's equations for the individual cavity}
&= \int \text{d} \mathbf{r} \mathbf{E}_{\Omega}^*(\mathbf{r})\cdot\epsilon(\mathbf{r})\mathbf{E}_{\Omega}(\mathbf{r} + \mathbf{R}) \\
&- \frac{1}{\left( \frac{\Omega_s}{c}\right)^2} \int \text{d}\mathbf{r} \nabla \times \nabla \times \mathbf{E}_{\Omega}^*(\mathbf{r} + \mathbf{R}) \cdot \mathbf{E}_{\Omega}(\mathbf{r}).
\intertext{Performing integration by parts}
&= \int \text{d} \mathbf{r} \mathbf{E}_{\Omega}^*(\mathbf{r}) \cdot \epsilon(\mathbf{r}) \mathbf{E}_{\Omega}(\mathbf{r}+\mathbf{R}) \\
&- \frac{1}{\left(\frac{\Omega_s}{c}\right)^2} \int \text{d} \mathbf{r} \mathbf{E}_{\Omega}^*(\mathbf{r}+\mathbf{R}) \cdot \nabla \times \nabla \times \mathbf{E}_{\Omega}(\mathbf{r}) \text{d} \mathbf{r}.
\intertext{By Maxwell's equations for the individual cavity}
&= \int \text{d} \mathbf{r} \mathbf{E}_{\Omega}^*(\mathbf{r}) \cdot \epsilon(\mathbf{r}) \mathbf{E}_{\Omega}(\mathbf{r}+\mathbf{R}) \\
&- \text{d} \mathbf{r} \mathbf{E}_{\Omega}^*(\mathbf{r}) \epsilon_{\Omega}(\mathbf{r}) \mathbf{E}_{\Omega}(\mathbf{r}+\mathbf{R}) \\
&=\kappa(-\mathbf{R}).
\end{align}






\begin{thebibliography}{99}
\bibitem{Serendipity}
"Serendipity" is a high performance computing cluster whose main features are 16 Dell Power edge R430 servers with each 2x Intel Xeon E5-2698 processors at 2.2GHz, with in total 640 cores and about 3.6 terabyte memory, connected through Infiniband. 
The head node is a Dell Power Edge R730.
\end{thebibliography}
\end{document}